%% file: template.tex
\documentclass[cameraready]{Interspeech}

\title{Quantizer-Aware Hierarchical Neural Codec Modeling for Speech Deepfake Detection}

\author[affiliation={1}]{Jinyang}{Wu}
\author[affiliation={1},correspondingauthor]{Zihan}{Pan}
\author[affiliation={2}]{Qiquan}{Zhang}
\author[affiliation={1}]{Sailor Hardik}{Bhupendra}
\author[affiliation={1}]{Soumik}{Mondal}

\address{
    $^1$ Agency for Science, Technology and Research (A*STAR), Singapore \\
    $^2$ The University of New South Wales, Australia 
}

\email{\{wu\_jinyang,panz\}@a-star.edu.sg}


\usepackage{comment}
\usepackage{amsmath,amssymb}
\usepackage{graphicx}
\usepackage{mathtools}
\usepackage{amsmath}
\usepackage{cite}

\begin{document}

\maketitle

\begin{abstract}
Neural audio codecs discretize speech via residual vector quantization (RVQ), forming a coarse-to-fine hierarchy across quantizers. While codec models have been explored for representation learning, their discrete structure remains underutilized in speech deepfake detection. In particular, different quantization levels capture complementary acoustic cues, where early quantizers encode coarse structure and later quantizers refine residual details that reveal synthesis artifacts. Existing systems either rely on continuous encoder features or ignore this quantizer-level hierarchy. We propose a hierarchy-aware representation learning framework that models quantizer-level contributions through learnable global weighting, enabling structured codec representations aligned with forensic cues. Keeping the speech encoder backbone frozen and updating only 4.4\% additional parameters, our method achieves relative EER reductions of 46.2\% on ASVspoof 2019 and 13.9\% on ASVspoof5 over strong baselines.

\end{abstract}

\section{Introduction}

Recent advances in text-to-speech (TTS) and voice conversion (VC) have made it increasingly easy to generate highly natural-sounding speech. Despite their perceptual realism, deepfake utterances often exhibit subtle low-level inconsistencies, such as imperfect transient modeling, over-smoothed spectral details, or locally distributed artifacts that are difficult to capture reliably under real-world distortions\cite{muller22_interspeech,yi2022add}. Designing detectors that are both sensitive to these fine-grained artifacts and robust across domains remains a central challenge.

A dominant trend in deepfake detection is to build detectors on top of self-supervised learning (SSL) speech encoders \cite{xuan2025wavesp,xuan2025fake,guragain2024speech}. SSL representations are strong and transferable, offering contextualized frame-level features that generalize well across tasks and domains. However, the same contextual abstraction that benefits semantic modeling may attenuate fine-grained local cues that are highly indicative of synthetic generation, especially when artifacts are weak, sparse, or partially masked by channel effects. This motivates exploring complementary representations beyond purely continuous SSL embeddings.

Neural audio codecs provide such a complementary view. Instead of producing only continuous features, modern codecs discretize speech through residual vector quantization (RVQ), forming a sequence of quantizers that progressively encode coarse-to-fine residual details~\cite{zeghidour2021soundstream,ding2025kimi}. This process induces a structured residual hierarchy: earlier quantizers capture dominant structure, while later ones encode finer residual information. From a forensic perspective, synthesis artifacts may not distribute uniformly across quantizers but concentrate in specific residual levels.

Despite this structural property, the potential of neural codec representations remains underexplored in speech deepfake detection. 
Existing codec-based representation learning studies, such as Codec2Vec \cite{tseng2025codec2vec}, primarily aim at learning general speech representations from discrete codec tokens, without considering their forensic utility in detection tasks. 
Meanwhile, recent detection frameworks \cite{yang2024codecdeepfakerobost} have begun incorporating pretrained codec models such as EnCodec \cite{defossez2022high} as additional feature extractors in multi-view architectures. 
However, in these approaches codec outputs are treated as generic speech representations, without explicitly modeling the hierarchical structure introduced by residual vector quantization (RVQ). Furthermore, while both SSL models and neural codecs provide powerful learned speech representations, their interaction in deepfake detection remains largely underexplored. 
In particular, how to effectively align codec representations with SSL-based encoders while preserving the quantizer-level hierarchy has not been systematically studied.




Motivated by this, we study lightweight SSL–codec integration with explicit quantizer-level modeling. We first establish strong and stable hierarchy-preserving baselines that keep codec residual levels disentangled and combine them with SSL features through late concatenation and a single projection layer. Building on this setup, we propose Quantizer-Aware Static Fusion (QAF-Static), a parameter-light operator that learns global importance weights over residual quantizers to form a hierarchy guided codec representation before integration with SSL features. This design injects an interpretable structural bias aligned with RVQ, while maintaining training stability and minimal computational overhead.

\section{Related Work}

\subsection{SSL-based deepfake detection and parameter-efficient adaptation}

Self-supervised learning speech encoders \cite{chen2021_wavlm,hsu2021_hubert} have become the dominant backbone for spoofing and deepfake detection due to their transferability and contextual abstraction. Prior studies reveal that spoofing cues are not uniformly distributed across transformer layers. Pan \textit{et al.}~\cite{pan2024attentive} propose attentive merging of hidden embeddings from pre-trained speech models, showing that earlier SSL layers can contain stronger anti-spoofing signals and enabling lightweight detectors with partial backbones. 

To further reduce adaptation cost, MoLEx~\cite{pan2025molex}  introduces a mixture-of-LoRA-experts framework that selectively adapts SSL encoders through parameter-efficient routing mechanisms. These works demonstrate that structured exploitation of SSL internals either across layers or experts can improve detection robustness without full fine-tuning.

Complementary to SSL-only systems, multi-view strategies have also been explored. For instance, combining SSL embeddings with spectral features (e.g., MFCC/LFCC/CQCC) via concatenation or attention improves generalization under unseen distortions~\cite{el2025two}. These approaches treat additional representations as complementary views but do not explicitly consider structural priors within those representations.







\subsection{Neural audio codecs: continuous latents and discrete RVQ hierarchy}

Neural audio codecs (NACs) compress waveforms into a latent space optimized for reconstruction~\cite{zeghidour2021soundstream,defossez2022high,xin2024speechtokenizer,defossez2022highfi}. 
Given an input waveform $x$, the encoder produces continuous latent features
\begin{equation}
z = E(x), \quad z \in \mathbb{R}^{T \times D},
\end{equation}
which are typically used directly as continuous representations in detection pipelines. To enable bitrate control and compact transmission, NACs further discretize $z$ via residual vector quantization (RVQ). RVQ approximates each latent vector $z_t$ through a sequence of residual codebooks:

\begin{align}
r^{(0)}_t &= z_t, \\
k_{t,q} &= \operatorname*{arg\,min}_k \left\lVert r^{(q-1)}_t - e_{q,k} \right\rVert_2^2, \\
r^{(q)}_t &= r^{(q-1)}_t - e_{q,k_{t,q}}.
\end{align}

for $q = 1, \dots, Q$, where $e_{q,k}$ denotes the $k$-th codeword in the $q$-th codebook. 
The reconstructed latent is given by
\begin{equation}
\hat{z}_t = \sum_{q=1}^{Q} e_{q,k_{t,q}}.
\label{eq:rvq_sum}
\end{equation}
\vspace{-0.1in}

This procedure induces a coarse-to-fine residual hierarchy \cite{zhang2023speechtokenizer}: earlier quantizers capture dominant structure, while later quantizers encode progressively finer residual details.
Yet this hierarchical organization is rarely modeled explicitly, as many approaches either rely on continuous latent representations or apply uniform aggregation over discrete codec embeddings.

In practice, codec representations have begun to appear in speech deepfake forensics, but are leveraged in different ways.
For example, multi-view detection systems incorporate pretrained neural codec models (e.g., EnCodec ) as additional feature extractors and fuse them with other acoustic or SSL-derived representations \cite{yang2024robust}; in these settings, the encoder's continuous outputs are typically used as features, rather than the discrete RVQ codes.
In a different direction, SafeEar\cite{li2024safeear} repurposes a codec-style RVQ stack into a decoupling model that separates semantic and acoustic tokens for content privacy, and performs detection using acoustic-only tokens.
Beyond detection, codec-aware forensics has also been explored for CodecFake source tracing \cite{chen2025codec} by predicting codec taxonomy attributes such as VQ type, auxiliary objectives, and decoder structures.

In contrast to treating codec outputs as generic features, partitioning RVQ layers for privacy, or using taxonomy labels for attribution, we focus on explicitly modeling the RVQ quantizer hierarchy itself and studying lightweight, hierarchy-aware SSL--codec fusion for deepfake detection.

\section{Method}
\label{sec:method}
\begin{figure}[!t] 
    \centering
    \caption{Overview of the proposed speech deepfake detection framework. 
SSL features from WavLM (with Attentive Merging) are fused with codec 
representations through quantizer-aware weighting over RVQ levels. 
A quantizer mean pooling baseline is included for comparison. }
    \includegraphics[scale=0.29, trim=2 70 2 100, clip]
     {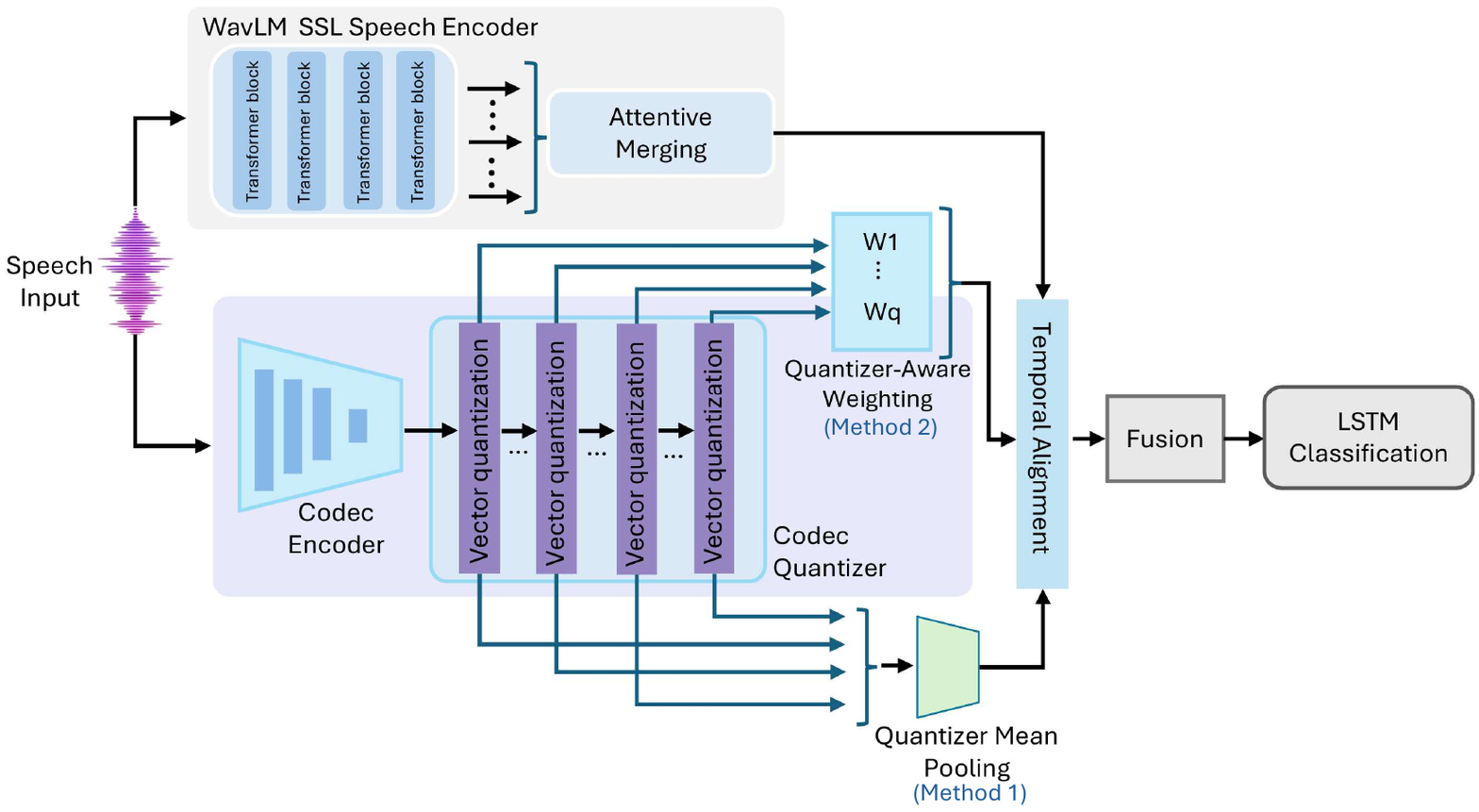}
    
    \label{fig:figure_pipeline}
\vspace{-0.2in}
\end{figure} 



Neural audio codecs represent speech through residual vector
quantization (RVQ), where successive codebooks progressively refine
the latent representation and form a coarse-to-fine hierarchy.
We hypothesize that this hierarchy is informative for deepfake
detection, since synthesis artifacts may distribute unevenly across
residual levels. However, existing approaches typically collapse
codec outputs into flat embeddings. To address this limitation, we introduce a lightweight
quantizer-aware operator that preserves RVQ structure and enables
hierarchy-aware aggregation of codec representations.




\subsection{Hierarchy codec modeling}

Residual vector quantization represents each latent vector
as a structured sum of residual levels (Eq.~\ref{eq:rvq_sum}),
naturally inducing an ordered hierarchy across quantizers.
Earlier levels encode coarse acoustic structure,
while later levels capture finer residual refinements.

A simple aggregation strategy applies uniform averaging across
quantizers (\textbf{Method~1}: Quantizer Mean Pooling):
\begin{equation}
\tilde{H}^{\mathrm{c}}_{\mathrm{avg}}
= \frac{1}{Q}\sum_{q=1}^{Q}\tilde{E}^{(q)}.
\end{equation}

However, this assumes equal importance across quantizers and
embedding dimensions, while synthesis artifacts may appear
unevenly across both hierarchical levels and feature channels.
To address this limitation, we introduce a lightweight
Quantizer-Aware Dimension-Wise Static Aggregation
(\textbf{Method~2}), which learns global importance weights across
residual quantizers while preserving the RVQ hierarchy






\subsection{Quantizer-Aware Dimension-wise Static Aggregation (QAF-Static)}

Neural codecs typically employ multiple residual codebooks to
progressively refine acoustic representations \cite{defossez2022_encodec}.
While standard practice aggregates codebook embeddings via
uniform averaging, such treatment implicitly assumes
equal contribution across quantization levels.
However, in practice, different codebooks capture
heterogeneous information \cite{zhang2023speechtokenizer}, and their relative importance
may vary across embedding dimensions \cite{zhang2023speak}. To enable fine-grained quantizer preference modeling
while preserving optimization stability,
we introduce a dimension-wise static aggregation mechanism.

Given $Q$ codec codebooks, each produces a frame-level embedding:
\begin{equation}
\tilde{E}^{(q)} \in \mathbb{R}^{B \times T \times D},
\end{equation}
where $D$ denotes the embedding dimension.

Instead of uniformly averaging or learning $Q$ scalar weights,
we introduce a dimension-wise static reweighting matrix:
\begin{equation}
\mathbf{W} \in \mathbb{R}^{Q \times D}.
\end{equation}

For each embedding dimension $d$, we normalize weights across codebooks:
\begin{equation}
\alpha_{q,d} =
\frac{\exp(W_{q,d}/\tau)}
{\sum_{q'=1}^{Q} \exp(W_{q',d}/\tau)},
\end{equation}
where $\tau$ is a temperature parameter that stabilizes training.

The aggregated codec representation is computed as:
\begin{equation}
\tilde{H}^{\mathrm{c}}_{b,t,d}
=
\sum_{q=1}^{Q}
\alpha_{q,d}
\,
\tilde{E}^{(q)}_{b,t,d}.
\end{equation}

This formulation enables each embedding channel to independently select
the most informative quantization group, providing fine-grained
quantizer-aware feature redistribution while maintaining
the stability of a static (input-independent) design. This mechanism can be interpreted as a channel-wise attention
over quantization groups, where each embedding dimension
performs a soft selection among codec codebooks.

\subsection{Lightweight SSL--Codec Fusion}
Once the codec hierarchy has been explicitly modeled,
we deliberately decouple structural aggregation from cross-stream interaction.
Our goal is to isolate the contribution of quantizer-aware modeling,
rather than attributing performance gains to increasingly complex
fusion modules. Therefore, we adopt a lightweight late-fusion strategy
that preserves representational independence between
the SSL contextual abstraction and the codec residual hierarchy.

After hierarchy-aware aggregation, we fuse the SSL and codec streams
via late concatenation followed by linear projection.

Given projected SSL features $H^{\mathrm{ssl}} \in \mathbb{R}^{B \times T \times d_{\text{model}}}$
and aggregated codec features
$\tilde{H}^{\mathrm{c}} \in \mathbb{R}^{B \times T \times d_{\text{codec}}}$,
we compute:

\begin{equation}
H^{\mathrm{f}} =
\mathrm{Linear}\big(
[H^{\mathrm{ssl}};\tilde{H}^{\mathrm{c}}]
\big).
\label{eq:late_concat}
\end{equation}

Late concatenation preserves representational independence
between continuous contextual abstraction (SSL)
and discrete residual hierarchy (RVQ).
This design avoids premature cross-stream interaction
and allows codec structure modeling to remain disentangled
from contextual abstraction.

We optionally apply Attentive Merging (AttM)~\cite{pan2024attentive}
to aggregate multi-layer SSL representations prior to fusion.
This component is orthogonal to RVQ hierarchy modeling
and serves as a backbone enhancement.

The fused representation is passed through a lightweight
single-layer LSTM followed by a linear classifier.
The decoder is intentionally compact to isolate
the contribution of representation modeling.

\section{Experimental Setup}

\subsection{Datasets and Evaluation Metric}

We evaluate our method on ASVspoof 2019 Logical Access (19LA) \cite{nautsch2021asvspoof}
and ASVspoof~5 \cite{wang2024asvspoof}.
For ASVspoof 2019 LA, we follow the official train/dev/eval protocol.
For ASVspoof~5, we use the standard train/dev/eval split.
Unless otherwise specified, models are trained separately on
19LA and ASVspoof~5 and evaluated on their respective evaluation sets.
Equal Error Rate (EER, \%) is used as the primary metric.

\subsection{Model Configuration and Training Details}

All systems employ WavLM-Large as the SSL backbone.
We use the first 12 transformer layers following the AttM configuration
\cite{pan2024attentive}. Unless AttM is enabled, the final layer
representation is used as $H^{\mathrm{ssl}}$.
For AttM, multi-layer SSL representations are aggregated as in
\cite{pan2024attentive}.
Importantly, the SSL backbone remains fully frozen during training.

For the codec stream, we adopt the Facebook EnCodec model
\cite{defossez2022highfi}, which employs residual vector quantization (RVQ).
The codec uses $Q{=}8$ quantizers, with a codebook size of 1024 and an
embedding dimension of 128.
Discrete codec indices are mapped to trainable embeddings before fusion.

All models are trained using Adam with identical optimization settings
across variants.
Early stopping is applied based on development-set EER to ensure fair
comparison.
Only the lightweight codec fusion layers and the final classifier are
optimized during training.

\input{Table/table_result1}
\section{Results and Discussion}
\label{sec:results}


\subsection{Hierarchy-aware modeling of codec representations}


Table~\ref{tab:table_result1} and Table 2 compare different fusion strategies on ASVspoof 2019 and ASVspoof~5. 
Our QAF-Static improves over the AttM baseline by 46\% relative EER
reduction on ASVspoof 2019 LA and achieves a further 13.9\% relative
improvement compared with the fully fine-tuned SSL model.

We first observe that codec-only modeling yields limited performance across datasets. 
While RVQ embeddings contain reconstruction-level cues, they lack the long-range contextual information provided by SSL features.
Adding a lightweight alignment module improves performance, indicating that matching the codec embedding space to the SSL feature scale is necessary.

We further analyze the parameter efficiency of the proposed codec-aware modeling.
The WavLM backbone contains approximately 315M parameters and is kept frozen throughout our training, while the neural codec branch introduces only about 14M additional parameters (approximately 4.4\% of the SSL backbone).
Despite this small parameter footprint, incorporating codec representations already improves performance over the AttM baseline even when the codec parameters remain frozen. 
When the codec branch is fine-tuned while keeping the SSL backbone frozen, the proposed model further surpasses strong SSL-based baselines that rely on full backbone fine-tuning, achieving state-of-the-art performance on ASVspoof 2019 LA and competitive results on ASVspoof5.

These observations suggest that codec representations provide complementary information beyond SSL features, particularly for capturing fine-grained signal artifacts that are important for deepfake detection.

\subsection{Complementarity with SSL-based representations}

We compare our hierarchy-aware codec modeling with Attentive Merging (AttM), which exploits the hierarchical structure of SSL transformer layers. 
While AttM models the layer-wise hierarchy within SSL representations, our approach focuses on the discrete residual hierarchy induced by RVQ. 
These two forms of hierarchy operate at different levels of representation.

\begin{figure}[!t] 
    \centering
    \caption{Learned quantizer contribution weights ($\alpha_q$) on the ASVspoof 5 dataset. Both the SSL encoder and codec encoder are frozen during training}
    \includegraphics[scale=0.33, trim=60 120 0 120, clip]
     {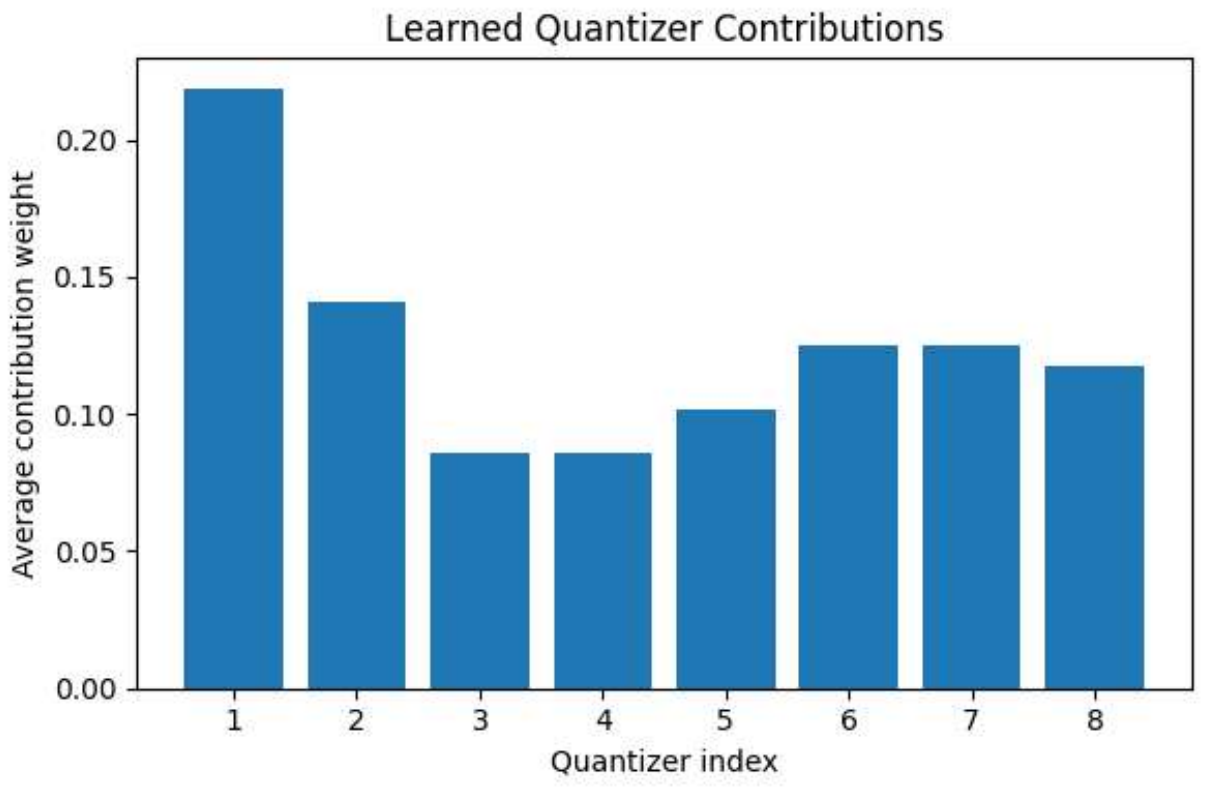}
    
    \label{fig:quantizer}
\vspace{-0.1in}
\end{figure}

Figure~\ref{fig:quantizer} illustrates the learned contributions of different codec quantizers. 
The distribution is clearly non-uniform, indicating that the model automatically learns the relative importance of quantizers instead of relying on uniform aggregation. 
Notably, the first quantizer receives the largest contribution, while intermediate quantizers receive smaller weights. 
This pattern reflects the hierarchical structure of residual vector quantization, where earlier quantizers capture coarse acoustic structure and later quantizers encode finer residual details.

Even with identical SSL backbones, incorporating quantizer-aware codec modeling provides complementary gains beyond SSL layer merging. 
This result suggests that RVQ hierarchy modeling captures forensic cues that are not fully represented by SSL features alone.



\subsection{Cross-codec robustness evaluation.}
\begin{figure}[!t] 
    \centering
    \caption{Detection performance across codec groups in the CodecFake benchmark.
The proposed quantizer-aware static fusion improves over the ATTM-LSTM baseline
and codec concatenation on codec family group B.}

    \includegraphics[scale=0.29, trim=0 240 0 170, clip]
     {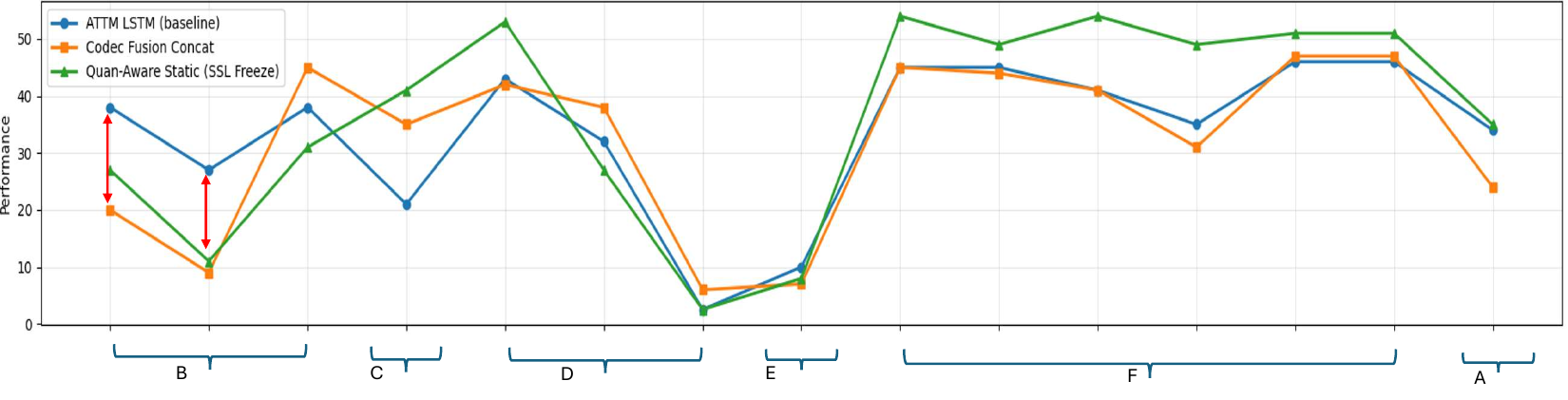}
    
    \label{fig:compare_codec}
\vspace{-0.1in}
\end{figure} 

To evaluate robustness against heterogeneous neural codec artifacts,
we adopt the CodecFake benchmark \cite{wu2024codecfake}, which comprises
multiple codec families with distinct quantization structures and bitrate
configurations. Different codecs employ diverse RVQ architectures and
compression strategies, resulting in varying distributions of quantization
artifacts.

Moreover, the most informative quantizer levels may vary across samples
and attack types. While our static quantizer-aware aggregation provides
a stable hierarchy prior, it may partially average out codec-specific
artifact cues when the artifact patterns differ significantly across
generation mechanisms. Therefore, evaluating cross-codec robustness is
important to verify whether the detector can generalize beyond the
training codec distribution.

In particular, Group B includes AcademiaCodec and HiFiCodec models,
which are characterized by relatively compact RVQ hierarchies and
lower effective bitrates, often producing more concentrated
quantization artifacts. This benchmark allows us to examine how the
detector generalizes across different codec generation mechanisms.
As shown in Fig.~\ref{fig:compare_codec}, the proposed fusion model
outperforms the ATTM-LSTM baseline on Group B (the first three subsets),
while yielding comparable performance on other codec families. These
results suggest that stronger and more consistent quantization artifacts
may provide clearer cues for codec-aware representation learning,
while generalization across diverse codec structures remains an open
challenge for future work.


\section{Conclusion}

\input{Table/table_results2}


We study codec-assisted speech deepfake detection from a
structured representation modeling perspective.
Instead of treating neural codec outputs as an unstructured auxiliary
stream, we explicitly model the residual hierarchy induced by RVQ.
We show that hierarchy-preserving fusion provides a strong baseline,
and propose Quantizer-Aware Static Fusion (QAF-Static), a lightweight
quantizer-weighting mechanism for codec aggregation.
Experiments demonstrate consistent improvements over strong SSL baselines. 
Notably, our approach keeps the SSL backbone frozen and introduces only about 4.4\% additional parameters relative to the backbone, showing that explicitly modeling the RVQ hierarchy provides an effective and parameter-efficient way to leverage neural codec representations for speech deepfake detection.

\bibliographystyle{IEEEtran}
\bibliography{mybib}

\end{document}

%% file: Table/table_result1.tex
\begin{table}[t]
\centering
\scriptsize
\setlength{\tabcolsep}{8.2pt}
\renewcommand{\arraystretch}{0.95}
\caption{Performance on ASVspoof5 Track 1 (EER \%, lower is better).
Relative improvement is computed with respect to the AttM baseline (6.60\%).
All our models keep the SSL encoder frozen during training.}
\label{tab:asvspoof5}
\vspace*{-0.05in}
\begin{tabular}{lcc}
\toprule
Method & EER (\%) & Rel.Impr. \\
\midrule
\multicolumn{3}{c}{\textit{Public baselines}} \\
\midrule
Wav2Vec2-AASIST-KAN \cite{jung2022_aasist} & 22.67 & -- \\
SEMAA-1 \cite{xia2024singleSEMAA} & 23.63 & -- \\
AASIST-CAM++ fused \cite{truong2024studyaasistcam} & 25.47 & -- \\
WavLM (FT-DA) \cite{combei2024wavlm} & 17.08 & -- \\
Best challenge submission \cite{wang2024asvspoof5} & 8.61 & -- \\
\midrule
\multicolumn{3}{c}{\textit{SSL-based baseline(SSL fine-tuned)}} \\
\midrule
ATTM + LSTM (baseline) \cite{pan2024attentive} & 6.60 & -- \\
\midrule
\multicolumn{3}{c}{\textit{Hierarchy-aware modeling (ours SSL Freeze)}} \\
\midrule
QAF-Static (codecF, sslF, Method 1) & 6.01 & 8.9\% \\
QAF-Static (codecF, sslF, Method 2) & 6.04 & 8.5\% \\
QAF-Static (codecT, sslF, Method 2) & \textbf{5.68} & \textbf{13.9\%} \\
\bottomrule
\end{tabular}
\label{tab:table_result1}
\vspace{-0.2in}
\end{table}

%% file: Table/table_results2.tex
\begin{table}[t]
\centering
\scriptsize
\setlength{\tabcolsep}{11.0pt}
\renewcommand{\arraystretch}{0.95}
\caption{Comparison on ASVspoof2019 LA (EER \%, lower is better).
Relative improvement is computed with respect to the AttM baseline (0.65\%).
codecF denotes the codec module is frozen, while codecT indicates the codec is fine-tuned.}
\begin{tabular}{lcc}
\toprule
Method & EER & Rel.Impr. \\
\midrule
\multicolumn{3}{c}{\textit{Public baselines}} \\
\midrule
W2V-XLSR-LLGF \cite{wang2021investigating}  & 2.80 & -- \\
HuBERT-XL \cite{wang2021investigating}  & 3.55 & -- \\
W2V-Large1-LLGF \cite{wang2021investigating}  & 0.86 & -- \\
W2V2-base-DARTS \cite{wang2022fully} & 1.19 & -- \\
W2V2-large-DARTS \cite{wang2022fully} & 1.08 & -- \\
\midrule
\multicolumn{3}{c}{\textit{SSL-based detectors}} \\
\midrule
AttM \cite{pan2024attentive} & 0.65 & -- \\
MoLEx \cite{pan2025molex} & 0.44 & 32.3\% \\
\midrule
\multicolumn{3}{c}{\textit{Hierarchy-aware modeling (ours)}} \\
\midrule
QAF-Static (codecF, sslF, Method 1) & 0.53 & 18.4\% \\
QAF-Static (codecF, sslF, Method 2) & 0.44 & 32.3\% \\
QAF-Static (codecT, sslF, Method 2) & \textbf{0.35} & \textbf{46.2\%} \\
\bottomrule
\end{tabular}
\label{tab:2}
\vspace{-0.2in}
\end{table}